
\def\mr {{\left (1-{M\over r}\right )}}
\def\qr {{\left (1-{Q_a^2\over Mr}\right )}}
\def\qm {{\left (1-{Q^2\over M^2}\right )}}
\def\htdz {{\delta H_{320}+{Q\over r^2}\chi_3}}
\def\derp {{\partial_+}}
\def\derm {{\partial_-}}

\input phyzzx.tex
\pubnum{ROM2F-92-36}
\titlepage
\title{ The vacuum polarization around}
\title{an axionic stringy black hole}
\author{ A. Carlini\foot{INFN, sezione di Genova, Via Dodecaneso 33, 16146
Genova, Italy} and A. Treves}
\address{SISSA, Strada Costiera 11, 34014 Trieste, Italy}
\author{ F. Fucito\foot{INFN,  sezione di Roma II, Via Carnevale, 00173 Roma,
Italy}}
\address{Dipartimento di Fisica, Universita' di Roma II ``Tor Vergata",
Via Carnevale, 00173, Roma, Italy}
\author{ M. Martellini\foot{INFN, sezione di Pavia, Via Bassi 6, 27100 Pavia,
Italy}}
\address{Dipartimento di Fisica, Universit\`a di Milano, Via Celoria 16,
20133 Milano, Italy}
\unnumberedchapters
\abstract
We consider the effect of vacuum polarization around the horizon of a
4 dimensional axionic stringy black hole.
In the extreme degenerate limit ($Q_a=M$),
the lower limit on the black hole mass for avoiding the polarization of the
surrounding medium is $M\gg (10^{-15}\div 10^{-11})m_p$ ($m_p$ is the proton
mass), according to
the assumed value of the axion mass ($m_a\simeq (10^{-3}\div 10^{-6})~eV$).
In this case, there are no upper bounds on the mass due to the absence of the
thermal radiation by the black hole.
In the nondegenerate (classically unstable) limit ($Q_a<M$), the black hole
always polarizes the surrounding vacuum, unless the effective cosmological
constant of the effective stringy action diverges.
\endpage
\pagenumber=1
If string theory is to describe a quantum theory of gravity, it is certainly
important to investigate what happens to it around black holes and cosmological
solutions and how such backgrounds are generated.
Recently low energy string theory solutions have been obtained in which
gravity is coupled to the Kalb-Ramond field \rlap.\Ref\kalb{M.J. Bowick, S.B.
Giddings, J.A. Harvey, G.T. Horowitz and A. Strominger\journal Phys. Rev. Lett.
&61(88)2823;\nextline
M.J. Bowick \journal Gen. Rel. Grav. &22(90)137;\nextline
B.A. Campbell, M.J. Duncan and K.A. Olive\journal Phys. Lett.& B251(90)34;
Phys. Lett. {\bf B263}(1991)364;
\nextline R.R. Hsu\journal Class. Quantum Grav.& 8(91)779;
\nextline R.R. Hsu and W.F. Lin\journal Class. Quantum Grav.& 8(91)L161.}
\Ref\raiten{E. Raiten, {\it Perturbations of a stringy black hole}, FERMI-PUB-
91/338-T, December 1991.}
In this context, one of the main problems concerns the stability of these
kinds of axionic black strings (ABS).
In a previous work\rlap,\Ref\martello{A. Carlini, F. Fucito and M. Martellini,
{\it On the stability of a stringy black hole}, {\it Phys. Lett.} {\bf B},
(1992), in press.}
it was shown that the ABS's are classically
and thermodynamically stable only in the extreme degenerate limit $Q_a=M$,
where
$Q_a~(M)$ is the axionic charge (mass).
However, it is well known that electrically charged black hole solutions of
General
Relativity, even in the degenerate case, may spontaneously loose their charge
because of vacuum polarization effects \rlap.\Ref\gibbons{G.W. Gibbons\journal
Comm. Math. Phys.&44(75)245.}

The purpose of this letter is to investigate under what circumstances
a degenerate 4-D ABS is stable against particle production from the
surrounding vacuum.
\REF\schwinger{J.S. Schwinger\journal Phys. Rev.&82(51)664.}
The method employed is a generalization of the effective action approach
of Ref.\schwinger ~developed for the semiclassical quantum electrodynamics.
Our unexpected result is that the degenerate ABS is `almost' stable for a wide
range of black hole masses.
The lower bound which we found crucially depends on the experimental
and astrophysical-cosmological estimates of the axion mass coupling.
We conclude the paper by performing a similar calculation for the
non-degenerate ABS solution ($Q_a<M$).
In this case the black hole always polarizes the surrounding vacuum,
loosing its axion charge.
This can be seen as the semiclassical counterpart of the classical
instability shown in Ref. \martello .

At the leading order in the string tension expansion $\alpha^{\prime}$,
the ABS solution is characterized by (see Ref. \martello ),
$$
\eqalign{&ds^2=-\mr dt^2+\qr dx^2+dy^2+{k~dr^2\over 8(r-M)
(r-Q_a^2/M)}, \cr}
\eqn\unoa
$$
$$
\eqalign{&H_{rtx}={Q_a\over r^2}, \cr}
\eqn\unob
$$
$$
\eqalign{&\Phi =\ln (r) +\Phi_{\infty}, \cr}
\eqn\unoc
$$
where $H\dot = dB$, $B$ is the Kalb-Ramond gauge field, $\Phi$ is the
dilaton and $\Phi_{\infty}$ its asymptotic value.

In this context, our strategy is to study the vacuum polarization phenomena by
treating the dilaton, the axion and the geometry in the external field
approximation.
In particular, we find the probability amplitude for the decay processes of the
axion field itself in a region close to the horizon, where we expect that the
vacuum polarization effects are dominant.
In the effective 4-D string theory, the leading decay process is
dictated by the coupling of the axion to the electromagnetic (quantum) field
$F_{\mu \nu}$, which appears in the order $\alpha^{\prime}$.
$F_{\mu \nu}$ is associated with a $U(1)$ subgroup of $E_8\times E_8$
($Spin(32)/Z_2$) of the superstring or coming from some kind of string
compactification.
At this string order, there are also vertices involving couplings
of the graviton and the Kalb-Ramond field\rlap,\Ref\campbell{B.A. Campbell,
N. Kaloper and K.A. Olive, {\it Classical hair for Kerr-Newman black holes in
string gravity}, CERN-TH-6332/91, December 1991} but in this semiclassical
approximation for gravity such couplings give only a `dressing' of the basic
ABS solutions (1-3).

Since we are interested in the vacuum polarization effects around the ABS
horizon at $r=M$, we assume for the dilaton the constant value:~~
$\Phi \simeq \Phi_{r=M} =\ln (M)\dot =\Phi_0$ (see eq. \unoc).
This turns out to imply that the $H$ equation of
motion at the leading order in $\alpha^{\prime}$ becomes
$\nabla_{\mu}H^{\mu \nu \lambda}=0$, and then one can write,
$$
H^{\mu \nu \lambda}=\epsilon^{\mu \nu \lambda
\rho}\nabla_{\rho}\theta ~~,\eqn\tre
$$
where $\theta$ is the axion pseudoscalar field.

Therefore, the low energy string action relevant for this problem has the form,
$$
S=\int d^4x\sqrt{-g}e^{\Phi_0}\left [ R+
{8\over k}+{1\over 2}(\nabla \theta)^2 +{\alpha^{\prime}\over 4}\left (
-F^2+\lambda\theta F~ ^{\ast} F\right )\right ]~~,~~\eqn\due
$$
where
$\lambda={m_a\over m_{\pi}
f_{\pi}}$, $m_a$ ($m_{\pi}$) is the axion (pion) mass, $f_{\pi}$ is the pion
decay constant, ~$8\over k$ is the
cosmological constant term corresponding to the central `charge deficit'
for the superstring\rlap,\Ref\dealwis{S.P. de Alwis, J. Polchinski and R.
Schimmrigk\journal Phys. Lett. &B218(89)449.}\foot{In the case of the
(SUSY) string coset-model $G/K$, where $G=SL(2, R)\times R \times
R$ and $K=U(1)$ of Ref. \raiten, one has that ($k=14/3$) ~$k=25/11$.}
and $^{\ast} F^{\mu \nu}={1\over 2} \epsilon_{~~\rho\sigma}^{\mu
\nu} F^{\rho\sigma}$, with $\epsilon_{~~\rho \sigma}^{\mu \nu}$ the covariant
Levi-Civita symbol.

\REF\coriano{C. Coriano'\journal Mod. Phys. Lett.&A7(92)1253.}
Following the Schwinger approach, the probability for the vacuum decay of
the axion into photons is,
$$
P_{\theta \rightarrow \gamma \gamma}\propto e^{-2Im~ \Gamma^1}~~,\eqn\quattro
$$
where $\Gamma^1$ is the one loop effective action coming from the external
field approximation (in $g_{\mu \nu}$ and $\theta$) of eq. \due , namely\rlap,
\foot{Eq. (7) can be obtained by the covariant generalization of
Ref.\coriano , after a rescaling of $H$ and $F$ by $e^{\Phi_0/2}$ and
having reabsorbed $\alpha^{\prime}$ in the definition of $F$.}
$$
\eqalign{2\Gamma^1=&\int ~d^4x\sqrt{-g}\biggl [-{\lambda^2\over
32\pi^2}\log (x_{\mu}x^{\mu}-i\epsilon)\biggl ({\lambda^2\over 4}\left
({H^2\over 6}\right )^2\cr
&-{1\over 12} H_{\mu \nu \sigma} \nabla^2 H^{\mu \nu \sigma}\biggr )
+{3\lambda^2\over 8\pi^2}{1\over (x^{\mu}x_{\mu}+i\epsilon)}
\left ({ H^2\over 6}\right )\biggr ]\cr}~~.\eqn\cinque
$$
Notice that, in eq. \cinque, in virtue of the ABS ansatz, we have
$H_{\mu \nu \lambda} \nabla^2 H^{\mu \nu \lambda}=0$.

In the extremal limit $Q_a=M$, where the ABS is classically stable, using
the classical background for $H^{\mu \nu \lambda}$ and $g_{\mu \nu}$
(eq. \unoa ~and \unob) and exploiting the
mathematical identities $\log (x_{\mu}x^{\mu}-i\epsilon)=\log (x^{\mu}x_{\mu})
-2\pi ni~~,~~n=0, 1, .....~$ and $~{1\over x^{\mu}x_{\mu}+i\epsilon}=P\left(
{1\over x^{\mu}x_{\mu}}\right )-\pi i\delta(x_{\mu}x^{\mu})$, one finds
(reintroducing dimensional factors of $M_p$, the Planck mass),
$$
\eqalign{2Im ~\Gamma^1=&\biggl [{3\lambda^2M^2\over \sqrt{8k}\pi M_p}\int
d^4x{1
\over r^3} \delta(g_{\mu \nu}x^{\mu}x^{\nu})\cr
&+{n\lambda^4M^4\over \sqrt{8k^3}\pi M_p}\int d^4x{1\over r^5}\biggr
]\biggr\vert_{r
\simeq M/M_p^2}~~.\cr}\eqn\sei
$$
We would like to stress the fact that one could have equally
well calculated the Schwinger effective amplitude for the vacuum polarization
by using a different conformally rescaled classical metric $\tilde g_{\mu\nu}
=e^{s\Phi}g_{\mu \nu}$, with $s$ arbitrary.
In this case, however, the new effective action \cinque ~ would correspond
to a different quantum theory\rlap,\Ref\birrel{N.D. Birrell and P.C.W. Davies,
{\it Quantum fields in curved space}, Cambridge University Press, Cambridge,
1982} and would not be trivially related to eq. \tre.

It is convenient to perform the integral over $t$ in the first term of the
right
hand side of \sei ~using the delta function expansion,
$$
\delta (g_{\mu \nu}x^{\mu}x^{\nu})=
{\delta (t+[1+\hat x^2+\hat y^2]
^{1/2}\mu)\over \left (1-{M\over M_p^2r}\right )\vert t\vert },\eqn\sette
$$
where we have defined,
$$
\eqalign{&\mu=\sqrt{{k\over 8}}{1\over M_p}\left (1-{M\over M_p^2r}\right )
^{-3/2}~~,\cr
&\hat x={x\over \mu}~~,\cr
&\hat y={y\over \mu}\left (1-{M\over M_p^2r}\right )^{-1/2}~~.\cr}\eqn\otto
$$
A reasonable estimate of the three-volume near the ABS horizon may be obtained
by putting the black hole in a box of linear dimension $M/M_p^2$.
In this ways one gets,
$$
\int dt\int dxdy \simeq l_{\theta}~{M^2\over M_p^4},\eqn\nove
$$
where $l_{\theta}$ is the Compton wavelength of the axion pseudoscalar.
Moreover, using \sette, \otto ~in the first term of the right hand side of \sei
{}~one can evaluate the double integral,
$$
\int^{\hat\beta} d\hat x\int^{\hat\gamma} d\hat
y[1+\hat x^2+\hat y^2]^{-1/2}\simeq \hat \beta \hat \gamma \bigr\vert_{r\simeq
M/M_p^2},\eqn\dieci
$$
which may be a fairly good approximation since,
$$
\eqalign{&\hat \beta\bigr\vert_{r\simeq M/M_p^2}={M\over M_p}\sqrt{8\over k}
\left (1-{M\over r}\right )^{3/2}\biggr\vert_{r\simeq M/M_p^2}\simeq 0~~,\cr
&\hat \gamma\bigr\vert_{r\simeq M/M_p^2}={M\over M_p}\sqrt{8\over k}\left
(1-{M\over r}\right )\biggr\vert_{r\simeq M/M_p^2}\simeq 0~~.\cr}\eqn\undici
$$
\REF\axion{{\it Review of particle properties}, {\it Phys. Rev.}
{\bf D45}(92)V.17.}
Therefore, reinserting back all relevant dimensional parameters
($m_{\pi}=135~~ MeV$ and $f_{\pi}=93 ~~MeV$).
Considering the following recent (astrophysical-cosmological) bounds for the
axion mass, $m_a=(10^{-3}\div 10^{-6})~~eV$  (see Ref. \axion ),
the final result for the Schwinger effective action is,
$$
\eqalign{2Im~ \Gamma^1 \simeq &{1\over k^{3/2}}\biggl [1.2~ (10^{95}\div
10^{89})
\left (1-{M\over M_p^2r}\right )^{1/2}\biggr\vert_{r\simeq M/ M_p^2}k^{1/2}\cr
&+6.2~(10^{143}\div 10^{134})\biggr ] \left ({M\over M_{\odot}}\right
)^2~~.\cr}
\eqn\dodici
$$

Now, the condition for avoiding the ABS discharge due to the vacuum
polarization can be directly read by \quattro , i.e. one should have
$2Im \Gamma^1 \gg 1$.
Then, from \dodici , discarding the first term, this condition implies a
lower bound for the ABS mass (for $n\not =0$),
$$
M\gg {k^{3/4}\over n^{1/2}}(1.6~ 10^{-15}\div 4.8 ~ 10^{-11})m_p~~,
\eqn\tredici
$$
where $m_p$ is the proton mass.
In the case of the extreme ABS solution, the temperature $T=0$ (see Ref.
\martello ), and therefore one does not expect a
thermal production of (virtual) particles around the horizon.
Therefore, the lower bound for $M$ which is given by \tredici ~is the only
relevant condition on the mass in order to avoid vacuum polarization of
the medium surrounding the extreme ABS solution.
The condition given by eq. \tredici ~ is by far much weaker than the usual
bounds on the general relativity black hole masses described in the
literature (see Ref. \gibbons ).

In the non-degenerate ($Q_a<M$) case of ABS, one can repeat similar
calculations starting from eq. \cinque, and it is quite straightforward to
obtain the following estimate for the one loop effective action,
$$
\eqalign{2Im \Gamma_{(Q_a<M)}^1\simeq &k^{-3/2}\biggl [7~(10^{20}\div 10^{14})
\left (1-{Q_a^2M_p^2\over M^2}\right )^{1/2}k^{1/2}\cr
&+8.4~(10^{-9}\div 10^{-18})~nQ_a^4~\left ({M_{\odot}\over M}\right )^2\biggr ]
{}~~.\cr}
\eqn\quattordici
$$
For this set of ABS solutions, the temperature $T={M_p\over \sqrt{2k}\pi}
\left (1-{Q_a^2M_p^2\over M^2}\right )^{1/2}$ is different from zero
(see Ref. \raiten), and one
must also take into account for the possible polarization of the vacuum by
`thermal' effects.
The condition for avoiding thermal polarization may be approximately
written as $T\ll 2m_a$, while the Schwinger probability for the `decay'
of the axion into photons is suppressed for $Im \Gamma_{(Q_a<M)}^1\gg 1$.
These two conditions combine to the following bound on the ABS mass,
$$
1.4k~(10^{-21}\div 10^{-15})\ll \left (1-{Q_a^2M_p^2\over M^2}\right )^{1/2}\ll
7\sqrt{k}(10^{-31}\div 10^{-35})~~.
\eqn\quindici
$$
This condition in general {\it cannot} be satisfied by any value of $M$
(unless $k \simeq 0$, which is equivalent to having an infinite effective
cosmological constant in the low energy stringy action \tre).
Therefore, the non degenerate ABS's seem to be unstable under classical
perturbations (see Ref. \martello), to polarize the vacuum
surrounding their horizon and to rapidly loose their initial axion charge.

\refout
\end

\vskip 2cm
\centerline{CASE $n=0$}
\vskip 2cm
$$
2\Im \Gamma^1 \simeq 7.4 \cdot 10^{156} \left ({M\over M_{\odot}}\right )^2
\left (1-{M\over M_p^2x_2}\right )\biggr \vert_{x_2\simeq {M\over M_p^2}}
\ll 3.7 \cdot 10^{129}\left ({M\over M_{\odot}}\right )^2
$$

\vskip 2cm
\centerline{DIRECT VACUUM POLARIZATION BY THE BLACK HOLE GRAVITATIONAL FIELD :}
\centerline{TO AVOID POLARIZATION, THE WORK PERFORMED BY THE G. F. MUST OBEY}
\vskip 2cm
$$
E_{grav}\sim {M_p^2\over M}< 2m_a
$$
\vskip 2cm
\centerline{CONDITION ON $M$ FOR AVOIDING DIRECT VACUUM DISCHARGE}
\vskip 2cm
$$
M\ge 6.5\cdot (10^{-8}-10^{-5})M_{\odot}
$$
\vskip 2cm
WHICH IS THEREFORE THE FINAL LIMIT FOR AVOIDING BOTH KIND OF POLARIZATIONS
\ref\coriano{C. Coriano'\journal Mod. Phys. Lett.&()}
\ref\frolov{A.I. Zel'nikov and V.P. Frolov, in {\it Quantum Gravity,
`Proceedings of the fifth seminar on quantum gravity', Moscow, 1986}
\end


If string theory is to describe a quantum theory of gravity, it is certainly
important to investigate what happens to it around singular backgrounds
\break and how
such backgrounds are generated. Addressing the first issue has led\rlap~
\Ref\sinback{G.Horowitz and A.Steif\journal Phys.Rev.Lett.&64(90)260;
Phys.Rev.Lett.{\bf 65}\break
(1990)\b1518; Phys.Lett.{\bf B258}(1991)91;\nextline
H.de Vega and Sanchez\journal Phys.Rev.Lett.&65(90)1517.}  to the
conclusion that in certain cases the propagation of strings through backgrounds
which are singular in the sense of classical gravity, is followed by the
excitation of an infinite number of modes of the string itself.
For what the second
problem is concerned, black hole type singularities were first obtained by
satisfying the equations of motion arising from the four dimensional low-energy
Lagrangian derived from string theory, which
describe the coupling of gravitational, dilatonic, Maxwell and antisymmetric
fields\rlap.\Ref\gibbon{G.W.Gibbons and K.Maeda\journal Nucl.Phys.&B298(88)741;
\nextline
D.Garfinkle, G.Horowitz and A.Strominger\journal Phys.Rev.&43D(91)3140;
\nextline C.G.Callan, R.C.Myers and M.J.Perry\journal Nucl.Phys.&B311(88/89)
673.}

Recently a new way to generate 2-D black hole backgrounds has been
devised\Ref\witten{E.Witten\journal Phys.Rev.&44D(91)314; \nextline
R.Dijkgraaf, H.Verlinde and E.Verlinde\journal Nucl.Phys.&B371(92)269.}
gauging a WZW model built on a coset manifold.

The black holes obtained in Ref.\gibbon~have been throughly analized in a
recent series of papers\rlap.\Ref\preskill{J.Preskill, P.Schwarz,
A.Shapere, S.Trivedi and F.Wilczek\journal Mod.Phys.Lett.&A6(91)2353;
\nextline A.Shapere, S.Trivedi and F.Wilczek, {\it Dual Dilaton Dyons},
IASSNS-HEP-91/33, June 1991;  \nextline
S.B.Giddings and A.Strominger, {\it Dynamics of Extremal Black Holes},
UCSB-TH-92-01, February 1992.}\Ref\wil{C.Holzhey and F.Wilczek, {\it Black
Holes as Elementary Particles}, IASSNS-HEP-91/71, December 1991.}
Quite surprisingly, the black holes obtained
in this fashion exhibit different thermodynamical properties from those
obtained from
general relativity. In some instances, for example, their scattering
behaviour is more similar to that of an elementary particle than a
thermal object.

\REF\gilbert{G.Gilbert, {\it On the Perturbations of String Theoretic Black
Holes}, UMDEPP 92-035, August 1991; {\it The Instability of String
Theoretic Black Holes}, UMDEPP 92-110, November 1991.}
\REF\raiten{E.Raiten, {\it Perturbations of a Stringy Black Hole},
FERMI-PUB-91/338-T.}
The purpose of this letter is to analyze the behaviour of a four
dimensional black hole obtained along the lines of Ref.\witten: we study
how it behaves under geometrical perturbations and we will briefly describe
its thermodynamical \break
properties which turn out (in the extremal case) to be different both
from those of black holes obtained from classical gravity and those of
Ref.\preskill. The study of perturbations of stringy black holes
in two and four dimensions has been
carried out in Ref.\gilbert, \wil, \raiten. Our results are
different from those of Ref.\raiten: {\it the black hole under study is stable
under perturbations of the metric
only in the extremal case}, thus supporting the conjecture that extremal
black holes might be stable ``quantum" ground state for the underlying theory.

\REF\horowitz{J.Horne and G.Horowitz, {\it Exact Black String Solutions in
Three Dimens-ions}, UCSBTH-91-39, July 1991.}
We now turn to the derivation of a four dimensional black hole
following Ref.\raiten, \horowitz.
The starting point of our analysis is the WZW action:
$$
L(g)={k Tr\over 4\pi}\left [\int_{\Sigma}d^2\sigma \left (g^{-1}\derp
gg^{-1}\derm g \right )-\int_B{d^3y \over 3}\left (g^{-1}dg\wedge g^{-1}dg
\wedge g^{-1}dg \right )\right ],
\eqn\uno
$$
where the integrals are over the three manifold $B$ and its boundary
$\Sigma$.
We now gauge a one dimensional subgroup $H$ of the symmetry group, with
action $g\rightarrow hgh^{-1}$, and introduce the gauge field $A_i$, whose
gauge transformations are:
$$
\cases{\delta a=2\epsilon a, &\cr
\delta b=-2\epsilon b, &\cr
\delta u=\delta v=0, &\cr
\delta x_i=2\epsilon c_i, &\cr
\delta A_i=-\partial_i \epsilon. &\cr}
\eqn\due
$$
The proposal of Ref. \raiten~ follows by adding two free bosons $x_1$ and $x_2$
to the
2-d black hole of Ref. \witten, that is by letting $G=SL(2, R)\times R \times
R$, and by modding out, besides the above $H$ subgroup, the translations in
both $x_1$ and $x_2$.
Parametrizing $SL(2, R)$ as:
$$
g=\pmatrix{a&u\cr
-v&b\cr},
\eqn\tre
$$
the gauged WZW action which is invariant under \due~becomes:
$$
\eqalign{L(g,A)=L(g)&+{k\over 2\pi}\int d^2\sigma A_+\left (b\derm a
-a\derm b -u\derm v +v\derm u +{4c_i\over k}\derm x_i\right )\cr
&+{k\over 2\pi}\int d^2\sigma A_-\left (b\derp a -a\derp b +u\derp v -v\derp
u +{4c_i\over k}\derp x_i\right )\cr
&+{2k\over \pi}\int d^2\sigma A_+A_-\left ( 1+{2c^2\over k}-uv\right ),\cr}
\eqn\quattro
$$
where a sum over $i=1, 2$ is assumed, and $c_1, c_2$ are constants such
that:
$$
c^2=c_1^2+c_2^2.
\eqn\pippa
$$
We then fix the gauge by setting $a=\pm b$, depending on the sign of
$1-uv$, and we choose to work in the ansatz:
$$
c_1=c_2={\sqrt {k}\over 2}\left ({M^2 \over Q^2}-1\right )^{-1/2}.
\eqn\sei
$$
By making the transformation of variables:
$$
\cases{u=e^{\sqrt {{2\over k}}\qm ^{1/2}t}\left ({r\over M}
-1\right )^{1/2}\qm ^{-1/2}&, \cr
v=-e^{-\sqrt {{2\over k}}\qm ^{1/2}t}\left ({r\over M}
-1\right )^{1/2}\qm ^{-1/2}&, \cr}
\eqn\sette
$$
and integrating out the gauge fields, the WZW action finally turns out:
$$
L={1\over \pi}\int ~d^2\sigma ~\left [g_{\mu \nu}\derm x^{\mu}\derp x^{\nu}~+
{1\over 2}B_{\mu \nu}(\derm x^{\mu}\derp x^{\nu}-\derp x^{\mu}\derm x^{\nu})
\right ],
\eqn\otto
$$
where $g_{\mu \nu}$ is the 4-d metric, $B_{\mu \nu}$ an antisymmetric
tensor field, and $x^{\mu}=(t, x_1, x_2, r)$.
It is then almost straightforward to show that requiring the fields to be an
extremum of the low energy effective action from string theory:
$$
S=\int ~d^4x\sqrt{-g}~e^{\Phi}\left [R+(\nabla \Phi)^2-{H^2\over 12}+{8\over k}
\right ],
\eqn\nove
$$
and redefining $x_1, x_2$ coordinates as:
$$
\cases{x_1={1\over \sqrt{2}}(x+y),\cr
x_2={1\over \sqrt{2}}(x-y),\cr}
\eqn\dieci
$$
the final form of the fields is:

The field equations coming from this effective action are\rlap:
\footnote{Our equation (16) contains a $H^2_{\mu\nu}$ term which is not
present in (4.3) of Ref.\raiten~and which will make the final equations
look quite different.}
$$
\eqalign{&\nabla_\lambda (e^{\Phi}H^{\lambda \mu \nu})=0, \cr}
\eqn\dodicia
$$
$$
\eqalign{&-{H^2\over 6}+\nabla^2\Phi +(\nabla \Phi)^2-{8\over k}=0, \cr}
\eqn\dodicib
$$
$$\eqalign{&
R_{\mu \nu}=\nabla_{\mu}\nabla_{\nu}\Phi+{1\over 2}g_{\mu \nu}\left (
\nabla^2\Phi+(\nabla \Phi)^2-{8\over k}-{H^2\over 6}\right )+{H_{\mu
\nu}^2\over 4}~\dot =~ T_{\mu \nu}, \cr}
\eqn\dodicic
$$
$$
\eqalign{&
dH=H_{\mu \nu \lambda ,\rho}-H_{\nu \lambda \rho ,\mu}+H_{\lambda \rho \mu,
\nu}-H_{\rho \mu \nu ,\lambda}=0. \cr}
\eqn\dodicid
$$

Let us now summarize the global structure of the above metric:
\item
\bullet
$Q>M$. In this regime our black hole exhibits neither a horizon nor,
contrary to the charged black holes of general relativity, a curvature
singularity (naked singularity).
\item \bullet
$Q<M$. The solution has a curvature singularity and two Killing horizons: an
outer horizon at $r=r_+=M$ and an inner horizon at $r=r_-={Q^2\over M}$.
Opposed to the general relativity black hole, the generator of time
translations remains space-like also for $r<r_-$. As a consequence the
manifold is time-like and light-like geodesically complete.
\item \bullet
$Q=M$. This is the extremal case in which $r_+=r_-$. With respect to the
general relativity solution, we notice that the metric is boosted
along the $x$ direction.

We now discuss the thermodynamics.
The temperature and entropy of our black hole are:
$$
T={1\over \pi M}\sqrt{{M^2-Q^2\over 2k}},
\eqn\tredicibis
$$
$$
S={A\over 4}\biggr \vert_{r_+}={\pi^2\over 2}\left (1-{r_-\over r_+}\right )
^{1/2},
\eqn\tredici
$$
where the coordinates $x, y$ are now periodic: $x\in [0, 2\pi ]$,
$y\in [0, \pi ]$ and $A$ is the horizon area.
We here remark the difference with the other stringy black hole of
Ref.\preskill~which has a temperature $T={1 \over 8\pi M}$, independent of
the charge.

In the extremal case the black hole under study has zero entropy and
temperature while the classical gravity (string) solution has zero (${1
\over 8\pi M}$) temperature and finite (zero) entropy\rlap.\footnote{The
authors of Ref.\gibbon~consider also a model with a parameter $a$ which
interpolates between the classical gravity case ($a=0$) and the string case
($a=1$). The thermodynamical properties of the black hole under study are
thus equivalent to the case $0<a<1$.}

Let us now investigate the range of validity of the thermal description of
the black hole defined by \undicia. According to Ref.\preskill:
$$
{\partial T\over \partial M}\biggr \vert_{Q}={Q^2\over \pi
\sqrt{2k}M^2(M^2-Q^2)^{1/2}}\gg 1,
\eqn\quattordici
$$
for the thermal description to break down. This is the case for the
extremal hole where $Q\longrightarrow M$. This is true independently of the
value of the mass similarly to what happens to the black hole of
Ref.\preskill~but in contrast with the extreme Reissner-Nordstr\"om
solution.\REF\birrel{N.D.Birrel and P.C.W.Davis, {\it Quantum Fields in
Curved Space}, (Cambridge University Press ,Cambridge, 1982).}
Following Ref.\birrel~we now discuss the domain of validity of the
semi-classical approximation. This approximation breaks down when:
$$
{1 \over M} {\partial M \over \partial t} \simeq T.
\eqn\birreldavis
$$
Using the Stefan-Boltzmann radiation law, this implies $T M \simeq A  T^4$.
In the extremal limit $T\mapsto 0$ and the previous formula is satisfied
independently of the value of the mass.

Let us now study the perturbations of the metric field. We rewrite the
metric as:
$$
g_{\mu \nu}=\pmatrix{-e^{2f_0}&0&0&0\cr
0&e^{2f_1}&0&0\cr
0&0&e^{2f_2}&0\cr
0&0&0&e^{2f_3}\cr },
\eqn\quindici
$$
and let us take as an {\it ansatz} for the axial and polar perturbations a
sufficiently general definition consistent with time-dependence and axial
simmetry:
$$
\delta g_{\mu \nu}=\delta g^A_{\mu \nu}+\delta g^P_{\mu \nu}=\pmatrix{-2\fa
e^{2f_0}&-\chi_0 e^{2f_1}&0&0\cr
-\chi_0 e^{2f_1}&2\fb e^{2f_1}&-\chi_2 e^{2f_1}&-\chi_3 e^{2f_1}\cr
0&-\chi_2 e^{2f_1}&2\fc e^{2f_2}&0\cr
0&-\chi_3 e^{2f_1}&0&2\fd e^{2f_3}\cr},
\eqn\sedicibis
$$
with $x^{\mu}=(t, x, r, y)$. In view of our choice, the first order
perturbations $\delta f_{\mu}, \chi_{\mu}$ are $x$ independent.

\REF\chandra{S.Chandrasekhar, {\it The Mathematical Theory of Black Holes},
(Clarendon Press, Oxford, 1983).}
We now have to compute the first order variation of the equation of motion
\dodicia-\dodicid. Following Ref.\chandra~we compute the variation of the
geometry and of the
energy-momentum tensor in the tetrad formalism. We thus rewrite the metric as:
$g_{\mu \nu}=e^{(a)}_{\mu}e^{(b)}_{\nu}\eta_{(a)(b)}$.
The variations of the tetrads are:
$$
\eqalign{&\delta e^{\mu}_{(0)}=(-\fa e^{-f_0},~\chi_0 e^{-f_0},~0,~0),\cr
&\delta e^{\mu}_{(1)}=(0,~~-\fb e^{-f_1},~~~~0,~~~~0),\cr
&\delta e^{\mu}_{(2)}=(0,~\chi_2 e^{-f_2},~-\fc e^{-f_2},~0),\cr
&\delta e^{\mu}_{(3)}=(0,~\chi_3 e^{-f_3},~0,~-\fd e^{-f_3}).\cr}
\eqn\sedici
$$
The variation of the components of the energy-momentum tensor are:
$$
\eqalign{&\delta T_{(1)(2)}=0, \cr}
\eqn\disettea
$$
$$
\eqalign{
&\delta T_{(1)(3)}={e^{-2f_2}\over 2r}\left [ e^{-2f_0-f_1}{Q\over r}\left (
\htdz \right )+e^{f_1}\chi_{23}\right ], \cr}
\eqn\disetteb
$$
$$
\eqalign{
&\delta T_{(0)(0)}=e^{-2f_0-2f_1-2f_2}\left [{Q\over r^2}\delta H_{012}-{Q^2
\over r^4}(\fa +\fb +\fc)\right ]+ \cr
&-e^{-2f_2}\biggl (f_{0,2}\fx_{,2}+ {\fa_{,2}\over r}\biggr )-\Delta, \cr}
\eqn\disettec
$$
$$
\eqalign{
&\delta T_{(1)(1)}={Q^2\over r^4}e^{-2f_0-2f_1-2f_2}(\fa +\fb +\fc)+e^{-2f_2}
\biggl (f_{1,2}\fx_{,2} + \cr
&+{\fb_{,2}\over r}\biggr )-{8Q\over k}\delta H_{012}+\Delta, \cr}
\eqn\disetted
$$
$$
\eqalign{
&\delta T_{(2)(2)}={Q^2\over r^4}e^{-2f_0-2f_1-2f_2}(\fa +\fb +\fc))+{e^{-2f_2}
\over r^2} [2\fc (1+rf_{2,2})+ \cr
&-\fc_{,2} r +r^2(\fx_{,2,2}-f_{2,2}\fx_{,2})]-{8Q\over k}\delta H_{012}
+\Delta, \cr}
\eqn\disettee
$$
$$
\eqalign{
&\delta T_{(3)(3)}={e^{-2f_2-2f_3}\over r}\fd_{,2}+e^{-2f_3}\fx_{,3,3}+\Delta,
 \cr}
\eqn\disettef
$$
$$
\eqalign{
&\delta T_{(0)(2)}=-{e^{-f_0-f_2}\over r}(\fc_{,0}+rf_{0,2}\fx_{,0}), \cr}
\eqn\disetteg
$$
$$
\eqalign{
&\delta T_{(0)(3)}={Q\over 2r^2}e^{-f_0-2f_1-2f_2}\delta H_{123}, \cr}
\eqn\disetteh
$$
$$
\eqalign{
&\delta T_{(2)(3)}=-{Q\over 2r^2}e^{-2f_0-2f_1-f_2}\delta H_{013}-{e^{-f_2}
\over r}\fc_{,3}, \cr}
\eqn\disettei
$$
where:
$$\eqalign{
&\Delta\dot=
-{1\over 2}e^{-2f_0}\fx_{,00}+{4r^2\over k}e^{2f_0+2f_1} \fx_{,22}+{1\over 2}
\fx_{,33}+{4r\over k}\biggl (3-{2M\over r}+ \cr
&-{2Q^2\over Mr}+{Q^2\over r^2}\biggr )\fx_{,2}+
{4r\over k}e^{2f_0+2f_1}(\fa +\fb -\fc +\fd )_{,2}-{8\over k}\fc + \cr
&-{8Q^2\over kr^2}(\fa +\fb)+{8Q\over k}\delta H_{012}. \cr}
\eqn\pippo
$$
After defining $\chi_{\alpha \beta}\dot =\chi_{\alpha, \beta}-
\chi_{\beta, \alpha}$, we are now ready to write all the perturbation
equations at first order:
$$
\eqalign{&\chi_{23,3}-e^{-2f_0}\chi_{20,0}=0, \cr}
\eqn\diciannovea
$$
$$
\eqalign{
&e^{-2f_1-f_0-f_2}\left [ \left ( e^{3f_1+f_0-f_2}\chi_{23}\right )_{,2}+
\left ( e^{3f_1-f_0+f_2}\chi_{30}\right )_{,0}\right ]= \cr
&=-{1\over r}\left [e^{-2f_0-f_1}{Q\over r}\left (\htdz \right )+
e^{f_1}\chi_{23}\right ]e^{-2f_2}, \cr}
\eqn\diciannoveb
$$
$$
\eqalign{
&e^{-2f_0}\fb_{,00}-{8r^2\over k}e^{2f_0+2f_1}\fb_{,22}-\fb_{,33}
+{4Q^2\over Mk}e^{2f_0}(\fc -\fd -\fa )_{,2}+ \cr
&-{8r\over k}\left (1+{Q^2\over 2Mr}-{3Q^2\over  2r^2}\right )
\fb_{,2}-{8Q^2\over kMr}\left (1-{2M\over r}\right )\fc=
{8Q\over k}\biggl [{Q\over r^2}(\fa+ \cr
&+\fb +\fc)-\delta H_{012}\biggr ]
+{4Q^2\over Mk}e^{2f_0}\fx_{,2}+{8r\over k}e^{2f_0+2f_1} \fb_{,2}, \cr}
\eqn\diciannovec
$$
$$
\eqalign{
&e^{-2f_0}\fc_{,00}-{8r^2\over k}e^{2f_0+2f_1}(\fa +\fb +\fd )_{,22}
-\fc_{,33}-{8r\over k}\biggl (1-{Q^2\over 2Mr}+ \cr
&+{M\over 2r}-{Q^2\over r^2}\biggr )\fa_{,2}
+{4M\over k}\left (1+{Q^2\over M^2}-{2Q^2\over Mr}\right )\fc_{,2}
-{4r\over k}\biggl (2-{M\over r}+ \cr
&-{Q^2\over Mr}\biggr )\fd_{,2}-{8r\over k}
\biggl (1-{M\over 2r}+{Q^2\over 2Mr}
-{Q^2\over r^2}\biggr )\fb_{,2}+{8M^2\over kr^2}\biggl (1+{Q^4\over M^4}+ \cr
&+{5Q^2\over M^2}-{r\over M}+{3Q^4\over M^2r^2}-{Q^2r\over M^3}
-{4Q^4\over M^3r}-{4Q^2\over Mr}\biggr )e^{-2f_0-2f_1}\fc
= \cr
&={8Q\over k}\biggl [{Q\over r^2}(\fa+\fb)+{1\over Q}\left (
{3Q^2\over r^2}-{M\over r}-{Q^2\over Mr}\right )\fc -\delta H_{012}\biggr ]+
\cr
&+ {8r\over k}e^{2f_0+2f_1}(r\fx_{,22}-\fc_{,2})
+{4r\over k}\left (2-{M\over r}-{Q^2\over Mr}\right )\fx_{,2}, \cr}
\eqn\diciannoved
$$
$$
\eqalign{
&-e^{-2f_0}(\fb +\fc +\fd )_{,00}+{8r^2\over k}e^{2f_0+2f_1}\fa_{,22}
+{8r\over k}\biggl (1+{M\over 2r}+ \cr
&-{3Q^2\over 2r^2}\biggr )\fa_{,2}
+\fa_{,33}+{4M\over k}e^{2f_1}(\fb -\fc +\fd )_{,2}
+{8M\over kr}\biggl (1- {M\over r}+ \cr
&-{2Q^2\over Mr}+{2Q^2\over
r^2}\biggr )e^{-2f_0}\fc = {8\over k}\biggl [Q \biggl [
\delta H_{012}-{Q\over r^2}(\fa
+\fb+\fc)\biggr ]+ \cr
&-re^{2f_0+2f_1} \biggl [{M\over 2r}e^{-2f_0}\fx_{,2}+\fa_{,2}\biggr ]
\biggr ], \cr}
\eqn\diciannovee
$$
$$
\eqalign{
&-{8r^2\over k}e^{2f_0+2f_1}\fd_{,22}-(\fa +\fb +\fc )_{,33}
-{8r\over k}\left (1-{Q^2\over r^2}\right )\fd_{,2}+ \cr
&+e^{-2f_0}\fd_{,00}={8r\over k}e^{2f_0+2f_1}\fd_{,2} +\fx_{,33}, \cr}
\eqn\diciannovef
$$
$$
\eqalign{
&-re^{f_1}(\fb +\fd )_{,02}+{1\over 2}{Q^2\over Mr}e^{-f_1}\fc_{,0}
+{M\over 2r}e^{-2f_0}\biggl [e^{f_1}\fd_{,0}+ \cr
&+\qm e^{-f_1}\fb_{,0}\biggr ]=-e^{f_1}\biggl [\fc_{,0}+{M\over 2r}e^{-2f_0}
\fx_{,0}\biggr ], \cr}
\eqn\diciannoveg
$$
$$
\eqalign{
&-e^{-f_0}(\fb +\fc )_{,03}={4Q\over k}e^{f_0}\delta H_{123}, \cr}
\eqn\diciannoveh
$$
$$
\eqalign{
&-e^{f_1}\left (r e^{f_0}(\fa +\fb )_{,23}+{1\over 2}{M\over r}e^{-f_0}
\fa_{,3}\right )+{1\over 2}e^{-f_1}\biggl [\biggl (1-{2Q^2\over Mr}+ \cr
&+{Q^2\over M^2}\biggr ){M\over r}e^{-f_0}\fc_{,3}
-{Q^2\over Mr}e^{f_0}\fb_{,3}\biggr ]= -e^{f_0+f_1}\fc_{,3}
-{Q\over 2r}e^{-f_0-f_1}\delta H_{013}, \cr}
\eqn\diciannovei
$$
$$
\eqalign{
&{k\over 8}e^{-2f_0-2f_1}\delta H_{013,3}+\biggl [r^2
\delta H_{012}+Q(\fd +\fx -\fa -\fb -\fc)\biggr ]_{,2}=0, \cr}
\eqn\diciannovel
$$
$$
\eqalign{
&e^{2f_0}\delta H_{123,3}-\left [\delta H_{012}+{Q\over
r^2}(\fd +\fx -\fa -\fb -\fc)\right ]_{,0}=0, \cr}
\eqn\diciannovem
$$
$$
\eqalign{
&\delta H_{013,0}+{8r\over k}e^{2f_0+2f_1}\left [re^{2f_0}\delta H_{123,2}+
\left (2-{M\over r}\right )\delta H_{123}\right ]=0, \cr}
\eqn\diciannoven
$$
$$
\eqalign{
&\left (\htdz \right )_{,0}=0, \cr}
\eqn\diciannoveo
$$
$$
\eqalign{
&\left (\htdz \right )_{,3}=0, \cr}
\eqn\diciannovep
$$
$$
\eqalign{
&\left (\htdz \right )_{,2}+{1\over r}\left (2-{Q^2\over Mr}\right )e^{-2f_1}
\left (\htdz \right )=0, \cr}
\eqn\diciannoveq
$$
$$
\eqalign{
&\delta H_{123,0}+\delta H_{013,2}-\delta H_{012,3}=0, \cr}
\eqn\diciannover
$$
$$
\Delta=0.
\eqn\diciannoves
$$
Following the ansatz \sedicibis~on the perturbations of the metric we
shall divide our equations \diciannovea-\diciannoves~into two sets which
we will call
``{\it axial}'' and ``{\it polar}''. The equations for the axial (polar)
perturbations will contain
only $\delta H_{1\alpha \beta}, \chi_0, \chi_2, \chi_3$ ($\delta H_{320}, \fa ,
\fb , \fc , \fd , \fx$).
The behaviour of the perturbations which is consistent with the simmetries of
our previous ansatz is:
$$
\cases{